\documentstyle[11pt,epsf]{article}
\begin{document}

\title{Comment on $\footnotesize < <$White-Noise-Induced Transport in Periodic
 Structures$\footnotesize >>$
 by J.\L uczka {\em et al.}}
\author{T. Hondou($^{\ast}$) and Y. Sawada($^{\ast \ast}$) \\
($^{\ast}$) Graduate School of Information Sciences, Tohoku University,
\\ Sendai, 980-77 Japan \\
($^{\ast \ast}$) Research Institute of Electrical Communication,
 Tohoku University
 \\  Sendai, 980-77  Japan}
\date{}
\maketitle



\begin{abstract}
 In the paper by J.\L uczka {\em et al.} ({\em Europhys. Lett.}, {\bf 31}
(1995) 431), the authors reported by rigorous calculation that an additive
 Poissonian white shot  noise can induce
a macroscopic current of a dissipative particle in a periodic potential
 -- even {\em in the absence} of
spatial asymmetry of the potential.
We argue that their main result is an obvious one caused by
  the spatially broken symmetry of a probability distribution of the
additive noise, unlike the similar result caused by chaotic noise
which has a symmetric probability distribution ({\em J.Phys.Soc.Jpn.},
{\bf  63} (1994) 2014).

\end{abstract}

\hspace{10mm}

In a recent letter \cite{White} in this journal, it was reported that a
dissipative particle in a
periodic potential can be transported macroscopically
by a white shot noise of
zero average, even {\em in the absence} of spatial asymmetry of the potential.
Namely, the following dynamics(eq.(1) and eq.(2) of \cite{White} respectively)
 was considered
\begin{equation}
 \dot{x}=-\frac{\partial V}{\partial x}+  \xi(t),
\label{eq:1}
\end{equation}
with
\begin{equation}
\xi (t) = \sum_{i=1}^{n(t)} y_{i} \delta (t-t_{i})
 - \lambda \langle y_{i} \rangle,
\end{equation}
where $V(x)$ is a periodic potential, $\xi(t)$ is a white shot noise, and
$n(t)$ is a Poisson counting process with a parameter $\lambda$ which
determines the average sojourn time between two $\delta-$kicks.
The positive-valued weights of the $\delta-$pulses $\{y_{i}\}$,
being independent of $n(t)$, are exponentially distributed
(eq.(3) of \cite{White}),
\begin{equation}
 \rho(y) = A^{-1} \exp [-y/A ], \, \, \, \, \, y \ge 0,
\end{equation}
with $\langle y_{i} \rangle \equiv A$.
 In this dynamics, the authors of ref.\cite{White} claim that they
obtained and analyzed a
unidirectional motion of the
particle for finite values of the parameters, $\lambda$ and $A$, and
found an analytical expression for a stationary current of a particle flow.

This subject, noise induced transport, is interesting, in principle,
as the authors discussed; however,
their main result is trivial from a physical insight, because
an existence of a net current is immediately
obtained before one performs rigorous calculation, but only
in terms of a spatial asymmetry of the additive noise.
Their claim
 that the first term
of the additive noise
 dominates the finite current (see, $r.h.s.$ of eq.(2))
 is
"{\em not} evident
{\em a priori}, because the $\delta$-spikes act over zero duration" is not
justified as one sees below.

 Let us restrict ourselves, for simplicity, in the case of the piecewise
linear potential (eq.(17) of ref.\cite{White}).
Then, one can find, at a glance, the condition
that a particle cannot
move at all in a left direction is
\begin{equation}
 | \frac{\partial V} {\partial x} | \ge \lambda \langle y_{i} \rangle,
\end{equation}
{\em i.e.} $1 \ge \lambda A \equiv a_{c}$.
In this condition, however, there exist a finite probability for a
particle to move in a right direction, as found in eq.(3):
{\em i.e.}
\begin{equation}
\int_{L/2}^{\infty} \rho (y) dy = \exp [-L/(2A)] > 0
\end{equation}
 gives a
 probability that a particle at one of the minima of the potential is
transferred to a next basin by a $\delta$-kick of an additive noise
at once (see eq.(17) of ref.\cite{White}), where $L$ is a period of the
potential.
This directly proves an existence of a unidirectional motion of the particle.
Therefore, it is evident {\em a priori}
that
the particle can be transported unidirectionally for $\lambda A \le 1$.
For a region $\lambda A > 1$, the way how a particle moves is not evident
{\em a priori}; however, an emergence of a net current is not surprising,
because the spatial symmetry of the probability distribution of the additive
noise is already broken: it would be rather surprising if the net current
disappeared at finite parameters, $\lambda$ and $A$.

 Finally, we
 remark that a particle can be transported unidirectionally in a
similar potential with spatial symmetry
 even if a probability density of an added
noise is {\em symmetric} and $\delta$ correlated\cite{Journal}.
 The unidirectional
motion can be attributed
 to a chaotic coherence\cite{Schuster} of which the effect is
caused by a $dynamical$ asymmetry.


\begin{thebibliography}{9}
\bibitem{White}
\L uczka J., Bartussek R. and H\"anggi P., {\em Europhys. Lett.}, {\bf 31}
(1995) 431.
\bibitem{Journal}
$a$) Hondou T., {\em J.Phys.Soc.Jpn.}, {\bf 63} (1994) 2014;
$b$) Hondou T. and Sawada Y., {\em Phys.Rev.Lett.} {\bf 75} (1995) 3269.
\bibitem{Schuster}
See, for example, Schuster H.G., {\em Deterministic Chaos} (Physik-Verlag,
Weinheim, 1984).
\end{thebibliography}
\end{document}